\documentclass[preprint,superscriptaddress]{revtex4}
\usepackage{graphicx}

\newcommand{\bra}[1] {\left\langle #1 \right|}
\newcommand{\ket}[1] {\left| #1 \right\rangle}

\begin{document}

\title{Controlling qubit-oscillator systems using linear parameter sweeps}

\author{Sahel Ashhab}
\author{Tomoko Fuse}
\affiliation{Advanced ICT Institute, National Institute of Information and Communications Technology, 4-2-1, Nukuikitamachi, Koganei, Tokyo 184-8795, Japan}

\author{Fumiki Yoshihara}
\affiliation{Advanced ICT Institute, National Institute of Information and Communications Technology, 4-2-1, Nukuikitamachi, Koganei, Tokyo 184-8795, Japan}
\affiliation{Department of Physics, Tokyo University of Science, 1-3 Kagurazaka, Shinjuku-ku, Tokyo 162-8601, Japan}

\author{Sunmi Kim}
\affiliation{Advanced ICT Institute, National Institute of Information and Communications Technology, 4-2-1, Nukuikitamachi, Koganei, Tokyo 184-8795, Japan}

\author{Kouichi Semba}
\affiliation{Advanced ICT Institute, National Institute of Information and Communications Technology, 4-2-1, Nukuikitamachi, Koganei, Tokyo 184-8795, Japan}
\affiliation{Institute for Photon Science and Technology, The University of Tokyo, 7-3-1 Hongo, Bunkyo-ku, Tokyo 113-0033, Japan}

\date{\today}

\begin{abstract}
We investigate the dynamics of a qubit-oscillator system under the influence of a linear sweep of system parameters. We consider two main cases. In the first case, we consider sweeping the parameters between the regime of a weakly correlated ground state and the regime of a strongly correlated ground state, a situation that can be viewed as a finite-duration quench between two phases of matter: the normal phase and the superradiant phase. Excitations are created as a result of this quench. We investigate the dependence of the excitation probabilities on the various parameters. We find a qualitative asymmetry in the dynamics between the cases of a normal-to-superradiant and superradiant-to-normal quench. The second case of parameter sweeps that we investigate is the problem of a Landau-Zener sweep in the qubit bias term for a qubit coupled to a harmonic oscillator. We analyze a theoretical formula based on the assumption that the dynamics can be decomposed into a sequence of independent Landau-Zener transitions. In addition to establishing the conditions of validity for the theoretical formula, we find that under suitable conditions, deterministic and robust multi-photon state preparation is possible in this system.
\end{abstract}

\maketitle

\section{Introduction}
\label{Sec:Introduction}

The study of cavity quantum electrodynamics (QED) has played a major role in shaping our understanding of various quantum phenomena and has led to the development of various new technologies \cite{Gerry,Walls,Scully}. In recent years the remarkable advances made with cavity QED systems utilizing superconducting circuits (now known as circuit QED) have accelerated the experimental realization of phenomena that had been predicted theoretically decades earlier but could not be observed in conventional cavity QED setups \cite{Blais}. The development of circuit QED has also spurred the emergence of new ideas and applications of cavity QED systems.

Some of the advantages of superconducting circuits in circuit QED systems are (1) the ability to reach previously inaccessible parameter regimes (e.g.~deep strong coupling), (2) the ability to set the circuit parameters by appropriate circuit design and (3) the ability to design circuits in which the parameters can be tuned in real time.

The tunability of circuit parameters is often utilized to choose optimal bias points and apply sinusoidal signals to drive transitions between different energy levels to manipulate the quantum state of the system. The tunability can also be utilized by performing parameter sweeps or quenches. While sinusoidal driving is studied more extensively, linear parameter sweeps are also important in quantum computing and related applications. For example, parameter sweeps can be used to perform quantum annealing and adiabatic quantum computing \cite{Hauke,Albash}. Understanding the quantum dynamics under such unidirectional parameter sweeps is crucial to designing good control protocols that utilize the parameter sweeps for efficient state preparation. One-way parameter variations in quantum systems can also lead to interesting phenomena such as Landau-Zener (LZ) transitions \cite{Shevchenko} and the creation of topological defects via the Kibble-Zurek mechanism \cite{DelCampo}. Having a good grasp on the probability that the system might leave the desired quantum state or the amount of energy generated as a result of the parameter sweep is crucial for practical applications. We will therefore investigate the manifestation of these effects in the present study.

One of the important tasks in circuit QED systems is the preparation of nonclassical states, such as Fock states and Schr\"odinger cat states. These could be useful for various applications, including quantum communication, sensing, and error correction. We will therefore pay special attention to the possibility of obtaining practically interesting states at the end of the parameter sweep. It is worth noting in this context that there are alternative proposals in the literature for generating nonclassical states in cavity QED systems using periodic modulation of the qubit-cavity coupling strength \cite{Xiao,Liu}. One potential advantage of linear sweeps is that they can be robust against certain fluctuations, depending on the details of the experimental setup under study.

In this work, we perform a systematic analysis of some practically relevant situations in which the parameters of a cavity QED system are varied in a linear sweep. We inspect the results to gain insight into the quantum system dynamics and to search for potential applications such as controlled state preparation. Two cases arise naturally when considering circuit QED systems with variable parameters. In the case where the qubit is biased at its symmetry point, the system possesses normal and superradiant phases, depending on the relation between the various system parameters. The normal phase can be thought of as the weak-coupling phase and is associated with weak qubit-cavity correlations in the low-lying energy eigenstates, while the superradiant phase is associated with strong coupling and strong qubit-cavity correlations in the low-lying energy eigenstates. We investigate the evolution of the system following a quench between the two different phases. We find in particular that the number and energy of the excitations created in the quench is qualitatively asymmetric between the cases of a normal-to-superradiant quench and a superradiant-to-normal quench. The other case in which a linear parameter sweep arises naturally in a circuit QED system is related to the LZ problem. In addition to causing a nonadiabatic transition in the qubit, the sweep can lead to the creation of photons in the cavity. We investigate the dynamics in this case and calculate the probabilities of various final outcomes. In particular, we show that a standard LZ sweep protocol can be used to almost deterministically prepare Fock states in the cavity.

The remainder of this manuscript is organized as follows: in Sec.~\ref{Sec:Rabi}, we introduce the Rabi model that describes the cavity QED systems of interest for the present study. In Sec.~\ref{Sec:Quench}, we describe our simulations and results for the problem of sweeping the system parameters across the normal-superradiant phase boundary. In Sec.~\ref{Sec:LZwO}, we consider the case of performing a linear LZ parameter sweep in a qubit coupled to a harmonic oscillator. In Sec.~\ref{Sec:Experiment} we discuss possible implementations using superconducting circuits. We summarize our conclusions in Sec.~\ref{Sec:Conclusion}.

\section{Rabi model}
\label{Sec:Rabi}

The quantum Rabi model (QRM) describes a qubit coupled to a harmonic oscillator \cite{Jaynes,Xie}. The QRM Hamiltonian can be expressed as
\begin{equation}
\hat{H} = - \frac{\Delta}{2} \hat{\sigma}_x - \frac{\epsilon}{2} \hat{\sigma}_z + \omega \hat{a}^{\dagger} \hat{a} + g \hat{\sigma}_z \left( \hat{a} + \hat{a}^{\dagger} \right),
\label{Eq:QRM}
\end{equation}
where $\Delta$ is the qubit gap, $\epsilon$ is the qubit bias parameter, $\omega$ is the oscillator frequency (which implies the convention $\hbar=1$), and $g$ is the qubit-oscillator coupling strength. The operators $\hat{\sigma}_{\beta}$ (with $\beta=x,y,z$) are the qubit's Pauli operators, while $\hat{a}$ and $\hat{a}^{\dagger}$ are, respectively, the oscillator's annihilation and creation operators.

Previous studies on the QRM generally focused on the symmetric case $\epsilon=0$, mainly because the physical systems that were modeled by the QRM in the past were typically most accurately described by the case $\epsilon=0$. In fact, it is common in the literature not to include the bias term in the QRM Hamiltonian. Recent technological advances, e.g.~with circuit QED systems using superconducting circuits, provided setups where $\epsilon$ can be finite and is usually easily tunable \cite{Ashhab2010}. We shall start by setting $\epsilon=0$ in the remainder of this section and in Sec.~\ref{Sec:Quench} but consider the case of varying $\epsilon$ in Sec.~\ref{Sec:LZwO}.

One reason why the case $\epsilon=0$ is of particular interest for the present study is that this case can exhibit the so-called superradiance phase transition. When $4g^2\ll\omega\Delta$ the system is in the normal phase, where the low-lying states are approximately given by
\begin{eqnarray}
\ket{\rightarrow, n} & = & \frac{1}{\sqrt{2}} \left( \ket{\uparrow} + \ket{\downarrow} \right) \otimes \ket{n}
\nonumber \\
\ket{\leftarrow, n} & = & \frac{1}{\sqrt{2}} \left( \ket{\uparrow} - \ket{\downarrow} \right) \otimes \ket{n},
\label{Eq:NormalES}
\end{eqnarray}
the qubit states are defined by $\hat{\sigma}_z\ket{\uparrow}=\ket{\uparrow}$ and $\hat{\sigma}_z\ket{\downarrow} = -\ket{\downarrow}$, such that $\hat{\sigma}_x\ket{\rightarrow} = \ket{\rightarrow}$ and $\hat{\sigma}_x\ket{\leftarrow} = - \ket{\leftarrow}$ with $\ket{\rightarrow}=\left( \ket{\uparrow} + \ket{\downarrow} \right) / \sqrt{2}$ and $\ket{\leftarrow}=\left( \ket{\uparrow} - \ket{\downarrow} \right) / \sqrt{2}$, and the quantum number $n$ is the number of excitations (or photons) in the oscillator ($\hat{n}=\hat{a}^{\dagger} \hat{a}$). The ground state is $\ket{\rightarrow, 0}$. When $4g^2\gg\omega\Delta$ the system is in the superradiant phase, where the low-lying states are approximately given by
\begin{eqnarray}
\ket{+, n} & = & \frac{1}{\sqrt{2}} \left( \ket{\uparrow}\otimes \hat{D}(-\alpha) \ket{n} + \ket{\downarrow}\otimes \hat{D}(\alpha) \ket{n} \right)
\nonumber \\
\ket{-, n} & = & \frac{1}{\sqrt{2}} \left( \ket{\uparrow}\otimes \hat{D}(-\alpha) \ket{n} - \ket{\downarrow}\otimes \hat{D}(\alpha) \ket{n} \right),
\label{Eq:SuperradiantES}
\end{eqnarray}
the displacement operator $\hat{D}(\alpha)$ is defined as $\hat{D}(\alpha)=\exp\left( \alpha \hat{a}^{\dagger} - \alpha^* \hat{a} \right)$, and $\alpha=g/\omega$. The ground state is $\ket{+, 0}$. In the semiclassical limit, i.e.~the limit in which treating the dynamical variables in the Hamiltonian as classical dynamical variables leads to good physical predictions, an abrupt change in behaviour (e.g.~in qubit-oscillator correlations) occurs at a certain point if one of the system parameters is varied. The semiclassical limit is typically associated with the Dicke model, in which a large number of qubits are coupled to the oscillator. In the case of a single qubit, a similar semiclassical limit is realized when $\omega/\Delta\to 0$ \cite{Bakemeier,Ashhab2013,Hwang}. In this case, an abrupt change occurs at the point $4g^2=\omega\Delta$, with qubit-oscillator correlations remaining negligibly small below this critical coupling strength but increasing to finite values above the critical point.

It will be useful for our analysis below to note that the energy eigenstates in Eqs.~(\ref{Eq:NormalES}) and (\ref{Eq:SuperradiantES}) are either symmetric or antisymmetric with respect to the parity operator $\hat{P}=\exp\left\{i\pi \left(\hat{a}^{\dagger} \hat{a} + \hat{\sigma}_+ \hat{\sigma}_- \right) \right\}$, where $\hat{\sigma}_{\pm}=\left( \hat{\sigma}_y \pm i \hat{\sigma}_z \right)/2$. (Note that this definition of $\hat{\sigma}_{\pm}$ is different from the standard one because the qubit term in our symmetric QRM Hamiltonian is proportional to $\hat{\sigma}_x$ and not $\hat{\sigma}_z$, which is more commonly used in the cavity-QED literature.) The Hilbert space can therefore be divided into parity-symmetric and parity-antisymmertic subspaces that are completely decoupled from each other. The parity-symmetric subspace contains the states $\left\{ \ket{\rightarrow,0}, \ket{\leftarrow,1}, \ket{\rightarrow,2}, \ket{\leftarrow, 3}, ... \right\}$, while the parity-antisymmetric subspace contains the states $\left\{ \ket{\leftarrow,0}, \ket{\rightarrow,1}, \ket{\leftarrow,2}, \ket{\rightarrow, 3}, ... \right\}$. Similarly, the parity-symmetric subspace contains the states $\left\{ \ket{+,0}, \ket{-,1}, \ket{+,2}, \ket{-, 3}, ... \right\}$, while the parity-antisymmetric subspace contains the states $\left\{ \ket{-,0}, \ket{+,1}, \ket{-,2}, \ket{+, 3}, ... \right\}$. This decoupling between the subspaces of the Hilbert space allows us to simplify the problem by focusing on only one of the subspaces in our calculations.

\section{Quenches between normal and superradiant phases}
\label{Sec:Quench}

One interesting situation that could be realized in a circuit QED system is a finite-duration quench in which one of the system parameters is swept through the critical point that separates the normal and superradiant phases. Since we normally think of correlations as resulting from the coupling term in the Hamiltonian, the most natural scenario to think of in this context is perhaps sweeping the coupling strength $g$, i.e.~setting $g=vt$ where $v$ is the sweep rate and $t$ is the time variable with initial value $t=0$ and final value $t=T$. The Hamiltonian can then be expressed as
\begin{equation}
\hat{H} = - \frac{\Delta}{2} \hat{\sigma}_x + \omega \hat{a}^{\dagger} \hat{a} + vt \hat{\sigma}_z \left( \hat{a} + \hat{a}^{\dagger} \right).
\end{equation}
At $t=0$ the qubit and oscillator are completely decoupled ($g=0$), meaning that the system starts in the normal phase. Provided that $vT\gg\sqrt{\Delta\omega}/2$, the system parameters correspond to the superradiant phase at the final time.

One complication with this scenario is that the energy eigenstates do not have $T$-independent asymptotic values, since $\alpha$ in Eq.~(\ref{Eq:SuperradiantES}) would increase indefinitely with $t$ if $g$ did so. In other words, physical quantities such as the number of photons in the system diverge as $t\to\infty$. We can in principle analyze quantities that converge to finite asymptotic values when $g\to\infty$. However, these quantities do not arise intuitively. We therefore take a different approach.

We fix $\omega$ and $g$, and we vary $\Delta$ to achieve the sweep between the normal and superradiant phases. It is worth mentioning here that tunable-$\Delta$ superconducting qubits have been demonstrated in a number of recent experiments \cite{Paauw,Zhu}. We first consider a sweep from a large value ($\Delta_i\gg 4g^2/\omega$; normal phase) at the initial time to a small value ($\Delta_f\ll 4g^2/\omega$; superradiant phase) at the final time:
\begin{equation}
\hat{H} = - \frac{\Delta_i - vt}{2} \hat{\sigma}_x + \omega \hat{a}^{\dagger} \hat{a} + g \hat{\sigma}_z \left( \hat{a} + \hat{a}^{\dagger} \right),
\end{equation}
with the final time given by $T=(\Delta_i-\Delta_f)/v\approx\Delta_i/v$. In the QRM, the asymptotic expressions for the lowest energy eigenstates in the limits $\Delta\rightarrow\infty$ and $\Delta\rightarrow 0$ are, to lowest order, independent of the exact values of $\Delta$. As a result, provided that $\Delta_i\gg 4g^2/\omega$ and $\Delta_f\ll 4g^2/\omega$, the results can be expected to be almost independent of the exact values of $\Delta_i$ and $\Delta_f$. As long as $g/\omega$ is finite, no physical divergences occur in the final state of the system.

It is worth noting here that a related problem was considered in Ref.~\cite{SaitoKeiji}. In that study, the authors took the initial and final values of $\Delta$ to be $\pm\infty$ and demonstrated that interesting states, e.g.~a single-photon state, can be generated at the end of the parameter sweep. However, it is difficult in realistic systems, e.g.~using superconducting circuits, to tune $\Delta$ through zero between positive and negative values \cite{Paauw,Zhu}. We therefore focus on the case where $\Delta$ does not change sign. A related experiment realized a quench between dynamically engineered effective quantum Rabi models with weak and strong interactions in a suitably rotating frame \cite{Zheng}.

It is also worth noting that the semiclassical limit, where there is a sharp boundary separating the normal and superradiant phases, is realized when $\Delta/\omega\gtrsim 10^3$ (see Ref.~\cite{Ashhab2013}). If we vary $\Delta$ between large and small values, the semiclassical limit will not be valid throughout the sweep. However, the key point here is that at the semiclassical phase transition point the system parameters obey the relation $\Delta/\omega=(2g/\omega)^2$. As a result, if $g/\omega$ is significantly larger than 1, then $\Delta/\omega$ will also be much larger than 1 when the point $\Delta/\omega=(2g/\omega)^2$ is crossed. In this case, the semiclassical picture is valid at the most crucial time during the parameter sweep, and an abrupt change is expected at this point during the sweep.

We take the initial state of the system to be the ground state, which is approximately given by $\ket{\rightarrow, 0}$ in Eq.~(\ref{Eq:NormalES}). This choice is motivated mainly by its experimental relevance, as the ground state is generally the easiest state to prepare. Since the ground state is symmetric with respect to the parity operator and we are considering the symmetric QRM ($\epsilon=0$), the parity is a conserved quantity, and the system remains in the parity-symmetric subspace throughout its evolution. We therefore confine our analysis below to this subspace of the Hilbert space, e.g.~when we refer to the first- or second-excited state, ignoring the parity-antisymmetric subspace of the Hilbert space.

In the adiabatic limit $v\to 0$, the system remains in the ground state of the instantaneous Hamiltonian, and hence ends up in the ground state of the Hamiltonian at the final time. In the fast-sweep limit $v\to\infty$, the total sweep time is infinitesimally short, and the system does not have time to experience any dynamical evolution. As a result, the final state probabilities are given by the overlaps between the initial state and final energy eigenstates. If we take the limits $\Delta_i\to\infty$ and $\Delta_f=0$, the initial and final energy eigenstates are given exactly by Eqs.~(\ref{Eq:NormalES}) and (\ref{Eq:SuperradiantES}), respectively. Hence, we can use the formula for the excitation-number probability distribution in a coherent state of a harmonic oscillator to obtain exact expressions for all the probabilities in the final state:
\begin{eqnarray}
P(n) & = & \left| \bra{n} \hat{D}(g/\omega) \ket{0} \right|^2
\nonumber \\
& = & e^{-\langle n \rangle} \frac{\langle n \rangle ^n}{n!},
\label{Eq:InitialFinalOverlap}
\end{eqnarray}
where $n$ stands for the $n$th excited state (with $n=0$ for the ground state), and $\langle n \rangle=(g/\omega)^2$. The function $P(n)$ peaks at $n = \langle n \rangle$.

To find the probabilities at intermediate values of $v$, we performed numerical simulations in which we solve the time-dependent Schr\"odinger equation. In the simulations, we generally set $\Delta_i/\omega=4\times 10^3$ and $\Delta_f/\omega=0$. We include up to $500$ states in the truncated Hilbert space. We divide the total time into $10^5$ time steps. To ensure that our results are not affected by the finiteness of these parameters, we vary $\Delta_i/\omega$, the size of the truncated Hilbert space and the number of time steps to verify that the results are essentially unchanged, as long as these parameters remain sufficiently large.

\begin{figure}[h]
\includegraphics[width=8.0cm]{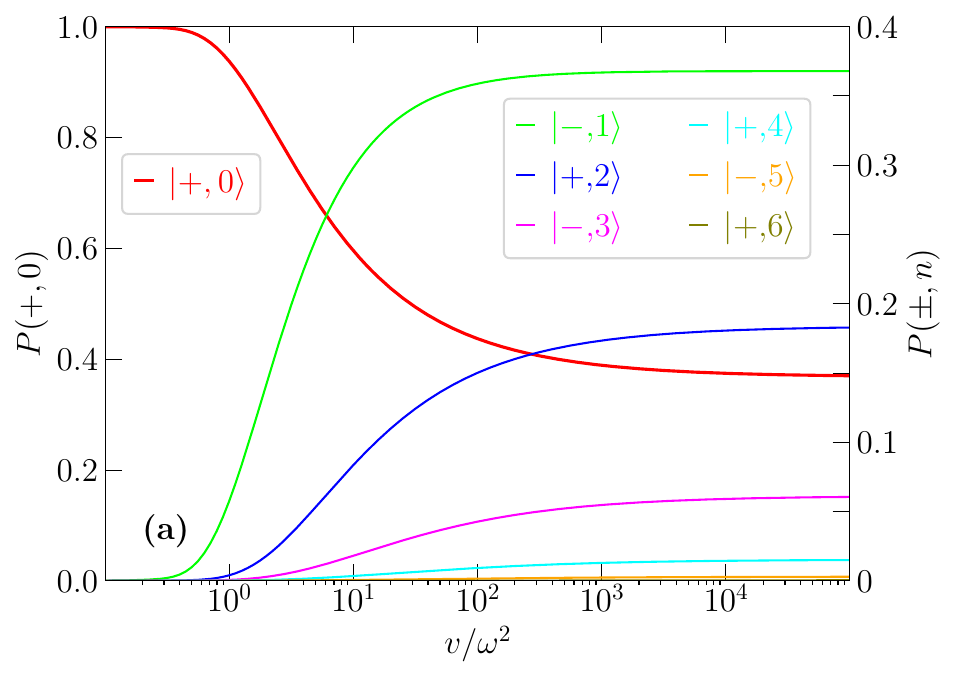}
\includegraphics[width=8.0cm]{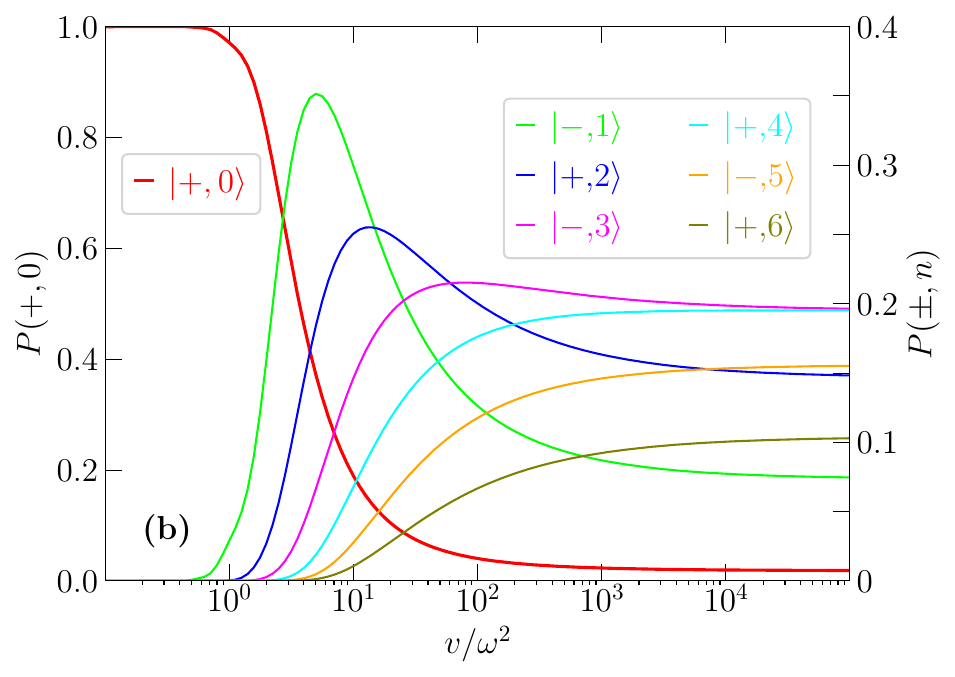}
\includegraphics[width=8.0cm]{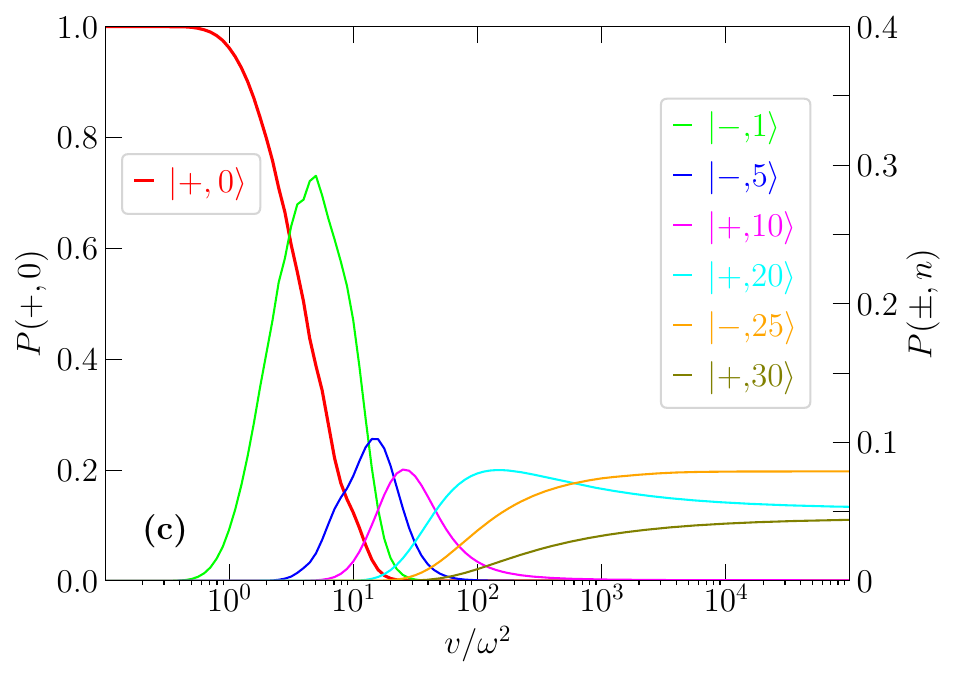}
\includegraphics[width=8.0cm]{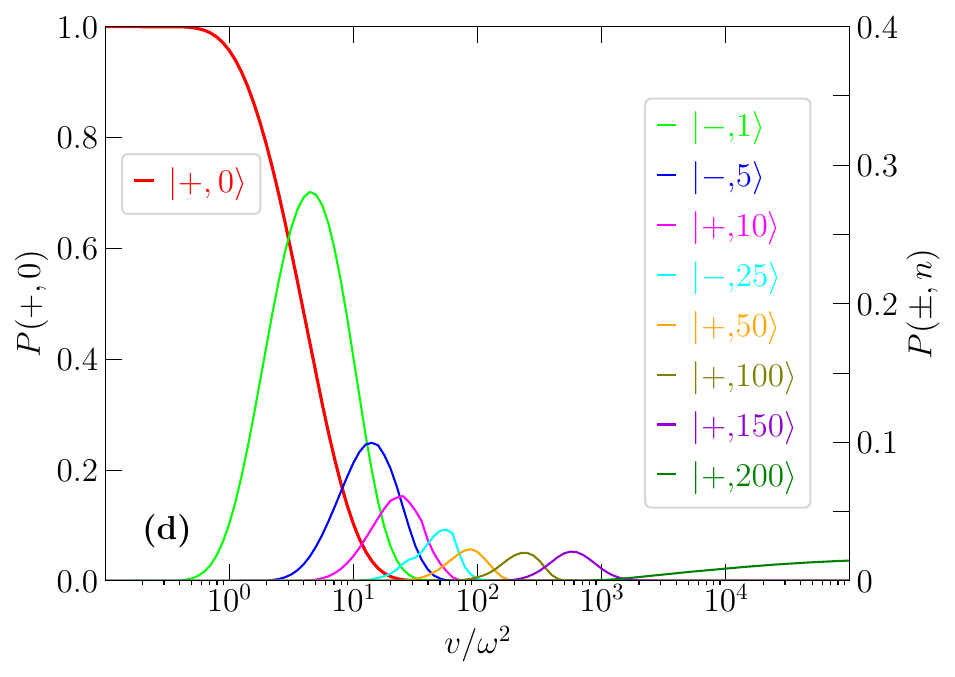}
\caption{Probabilities $P(\pm,n)$ as functions of $v/\omega^2$ for the QRM following a sweep of $\Delta/\omega$ from $4\times 10^3$ to zero, i.e.~from the weakly-correlated to the strongly-correlated regime. The different lines correspond to different final states, as explained in the legends. The y axis for $P(+,0)$ is on the left-hand side of each frame, while the y axis for all other lines is on the right-hand side of the frame. In Panels (a), (b), (c) and (d), $g/\omega=1$, 2, 5 and 20, respectively.}
\label{Fig:RabiQuenchProbabilitiesNS}
\end{figure}

The occupation probabilities $P(\pm,n)$ of a few representative final states (in the parity-symmetric part of the Hilbert space) as functions of the sweep rate (in the dimensionless combination $v/\omega^2$) for a few different values of $g/\omega$ are plotted in Fig.~\ref{Fig:RabiQuenchProbabilitiesNS}. As expected, the system remains in the ground state in the adiabatic limit in all cases. The probabilities change as the sweep rate increases until they reach the asymptotic values given in Eq.~(\ref{Eq:InitialFinalOverlap}) in the fast-sweep limit. For relatively small values of $g/\omega$ (Fig.~\ref{Fig:RabiQuenchProbabilitiesNS}a), all the probabilities $P(\pm,n)$ evolve monotonically between the two limits $v/\omega^2\to 0$ and $v/\omega^2\to\infty$. For large values of $g/\omega$ (Figs.~\ref{Fig:RabiQuenchProbabilitiesNS}b-d), where the highest occupation in the fast-sweep limit corresponds to a large value of $n$, intermediate peaks occur in $P(\pm,n)$ when plotted as functions of $v/\omega^2$. First $P(-,1)$ has a peak, then $P(+,2)$ and so on, until $n$ reaches $\langle n \rangle$. The probabilities $P(\pm,n)$ for $n\gtrsim\langle n \rangle$ increase monotonically with increasing $v/\omega^2$ and reach their asymptotic values at $v/\omega^2\to\infty$.

One overall feature that we can see in Fig.~\ref{Fig:RabiQuenchProbabilitiesNS} is that the maximum values of $P(\pm,n)$, i.e.~the heights of the $P(\pm,n)$ peaks, generally decrease with increasing $n$. Although it is not clear in the plots shown in Fig.~\ref{Fig:RabiQuenchProbabilitiesNS}, this trend is not monotonic: after a clear decrease for small values of $n$, the peak heights stabilize around a certain value (with a weak oscillatory behaviour). For $n>\left\langle n \right\rangle$, where there are no peaks anymore, the maximum values of $P(\pm,n)$ [at $v/\omega^2\to\infty$] decrease with increasing $n$ until they vanish in the limit $n\to\infty$.

There are a few additional small but interesting features in the behaviour of $P(\pm,n)$. There is a weak wavy behaviour in the probabilities $P(\pm,n)$ as functions of $v/\omega^2$. The wavy behaviour is more clearly visible at higher values of $g/\omega$. By comparing the lines for different values of $g/\omega$ (not shown in a single plot in the figure), we find that in the adiabatic regime (around $v/\omega^2\sim 1$) $P(+,0)$ exhibits non-monotonic behaviour as a function of $g/\omega$.

Considering the $g/\omega$ values used in the different panels in Fig.~\ref{Fig:RabiQuenchProbabilitiesNS}, Panel (c) has $\Delta/\omega=(2g/\omega)^2=100$ and Panel (d) has $\Delta/\omega=(2g/\omega)^2=1600$ at the semiclassical crossing point, which are therefore expected to be at least close to the semiclassical regime \cite{Ashhab2013}. However, the probabilities $P(\pm,n)$ do not exhibit any drastically different behaviour that indicates the crossing of a phase transition point, except for the fact that a large number of photons are generated by a fast sweep. This result is rather surprising, since one would expect a qualitatively different behaviour associated with the crossing of a sharp phase transition boundary, compared to cases where no such sharp boundary exists.

\begin{figure}[h]
\includegraphics[width=8.0cm]{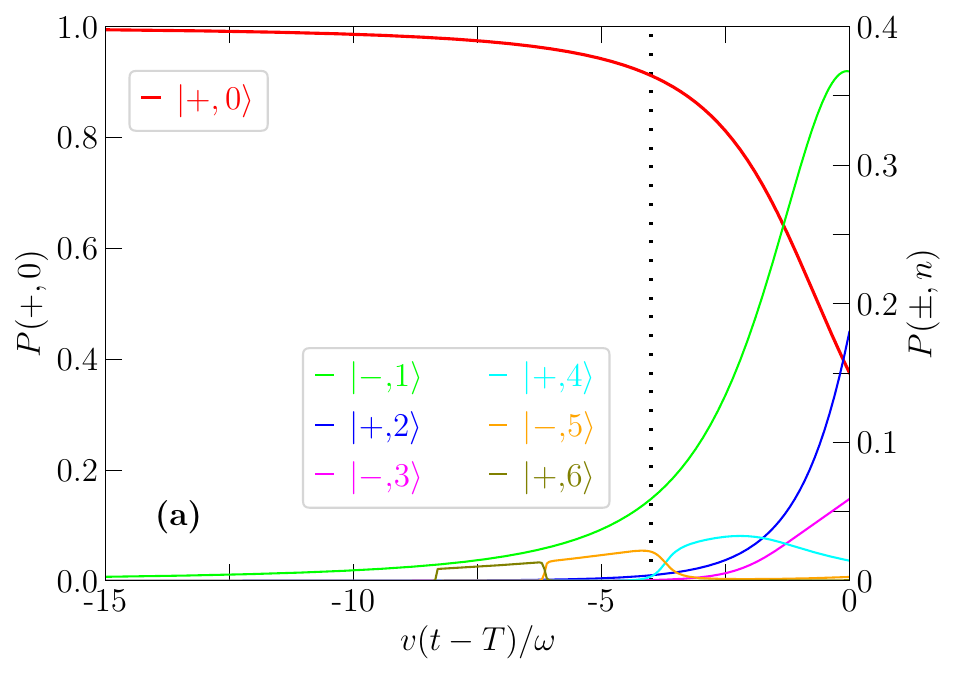}
\includegraphics[width=8.0cm]{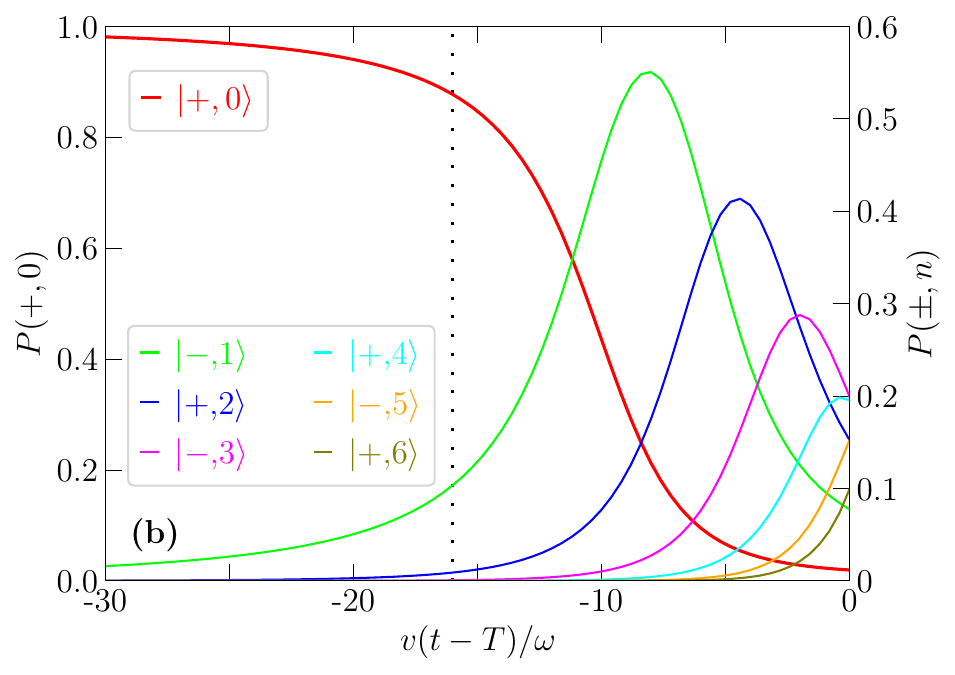}
\includegraphics[width=8.0cm]{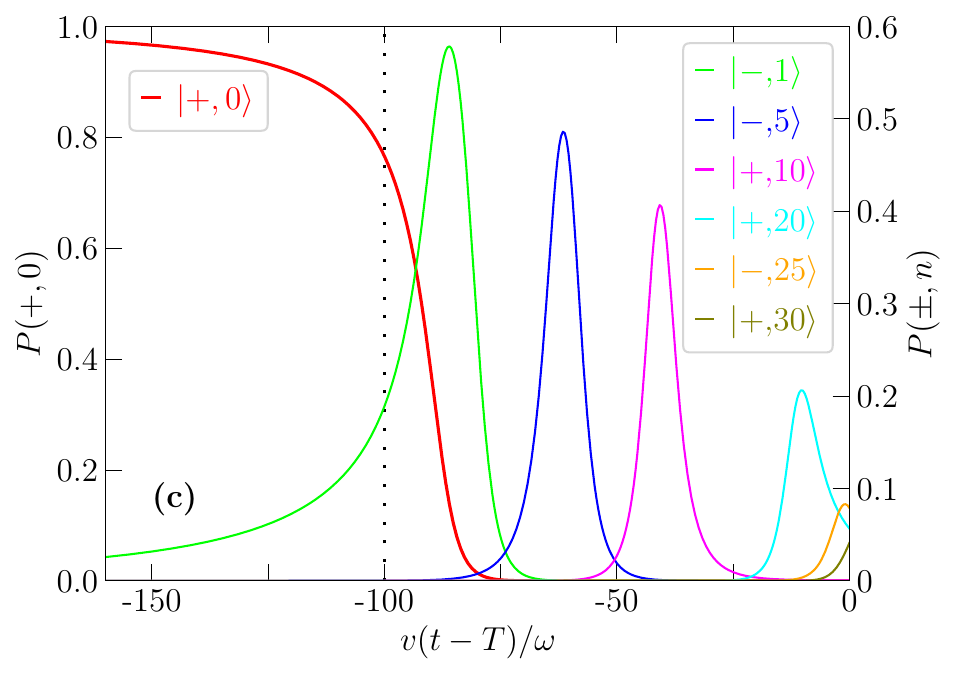}
\includegraphics[width=8.0cm]{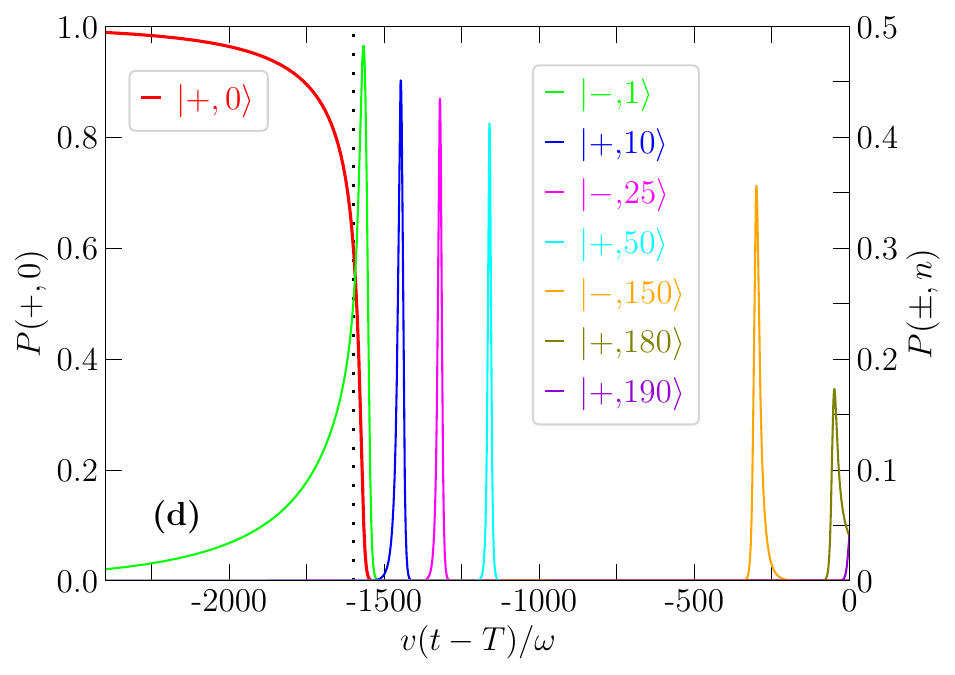}
\caption{Probabilities $P(\pm,n)$ as functions of time $t$ (measured relative to the final time $T$) for a weak-to-strong-correlation sweep. Each panel in this figure has the same value of $g/\omega$ as the panel with the same label in Fig.~\ref{Fig:RabiQuenchProbabilitiesNS}. In all cases, we set $v/\omega^2=10^4$. The vertical dotted lines mark the semiclassical phase transition crossing point.}
\label{Fig:RabiQuenchPopulationDynamicsNS}
\end{figure}

To gain further insight into the dynamics, we plot the probabilities $P(\pm,n)$ as functions of time $t$ for the fast-sweep regime in Fig.~\ref{Fig:RabiQuenchPopulationDynamicsNS}. In all cases, the system remains mostly in the ground state until it reaches the semiclassical crossing point, which is given by $v(t-T)/\omega=-(2g/\omega)^2$. Then state mixing starts and continues until the final time. In the case of relatively small $g/\omega$ (Fig.~\ref{Fig:RabiQuenchPopulationDynamicsNS}a), one interesting feature is that somewhat highly excited states are temporarily populated at intermediate times, even though only low-lying states are populated at the final time. Since we are dealing with a dynamic situation, and the direct nonadiabatic transitions to high energy levels can be stronger than those between neighbouring energy levels, this situation is not particularly surprising. For stronger coupling (Fig.~\ref{Fig:RabiQuenchPopulationDynamicsNS}b-d) a peak in $P(-,1)$ appears first, followed by a peak in $P(+,2)$ and so on, until the final time. As $g/\omega$ increases, the drop in $P(+,0)$ becomes increasingly sudden. This behaviour is consistent with what is expected in the semiclassical limit: when the transition point is crossed, excitations are created in analogy with the Kibble-Zurek mechanism. As can be seen most clearly in Fig.~\ref{Fig:RabiQuenchProbabilitiesNS}d, the fast drop in $P(+,0)$ at $v(t-T)/\omega=-(2g/\omega)^2$ is accompanied by a fast rise in probabilities $P(\pm,n)$ with small values of $n$. The state of the system then keeps moving towards higher values of $n$ until the final time. In other words, photons keep being added to the system until the final time. The fact that the semiclassical-picture dynamics occurs only for $(2g/\omega)^2 \gtrsim 10^3$ is consistent with the results of Ref.~\cite{Ashhab2013}, where it was shown that the sharp boundary between the normal and superradiant phases requires that $\Delta/\omega\gtrsim 10^3$. One implication of the rapid succession of peaks in the case of large $g/\omega$ (Fig.~\ref{Fig:RabiQuenchPopulationDynamicsNS}d) is that the final state is highly sensitive to the value of $\Delta_f$. In other words, if $\Delta_f$ is small but finite, the final state will have a large number of photons but will have small overlap with the final state for $\Delta_f=0$.

Since we are interested in the question of controlling the system via linear parameter sweeps, we look at the possibility of preparing specific states by a proper choice of the sweep rate. Apart from $P(+,0)$, which approaches 1 in the adiabatic limit, no other probability reaches a large value (all below 0.4 in Fig.~\ref{Fig:RabiQuenchProbabilitiesNS}). For example, the highest possible value of $P(-,1)$, which is given by 0.368, is obtained in the fast-sweep limit ($v/\omega^2\to\infty$) at $g/\omega=1$. For smaller or larger values of $g/\omega$, the maximum value of $P(-,1)$ remains below 0.368, and it vanishes in the limits $g/\omega=0$ and $g/\omega\to\infty$. The maximum value of $P(+,2)$ at $g/\omega=1$ (Fig.~\ref{Fig:RabiQuenchProbabilitiesNS}a) is 0.184. When $g/\omega=\sqrt{2}$, which is the point at which Eq.~(\ref{Eq:InitialFinalOverlap}) peaks at $n=2$, the maximum value of $P(+,2)$ is $e^{-2} \times 2^2/2!=0.271$. When $g/\omega=2$ (Fig.~\ref{Fig:RabiQuenchProbabilitiesNS}b), the maximum value of $P(+,2)$ is 0.255. Similarly to $P(-,1)$, and all $P(\pm,n)$ with $n\neq 0$, $P(+,2)$ vanishes in the limits $g/\omega=0$ and $g/\omega\to\infty$. These results show that it is not possible to prepare specific energy eigenstates with a high probability using this approach.

\begin{figure}[h]
\includegraphics[width=8.0cm]{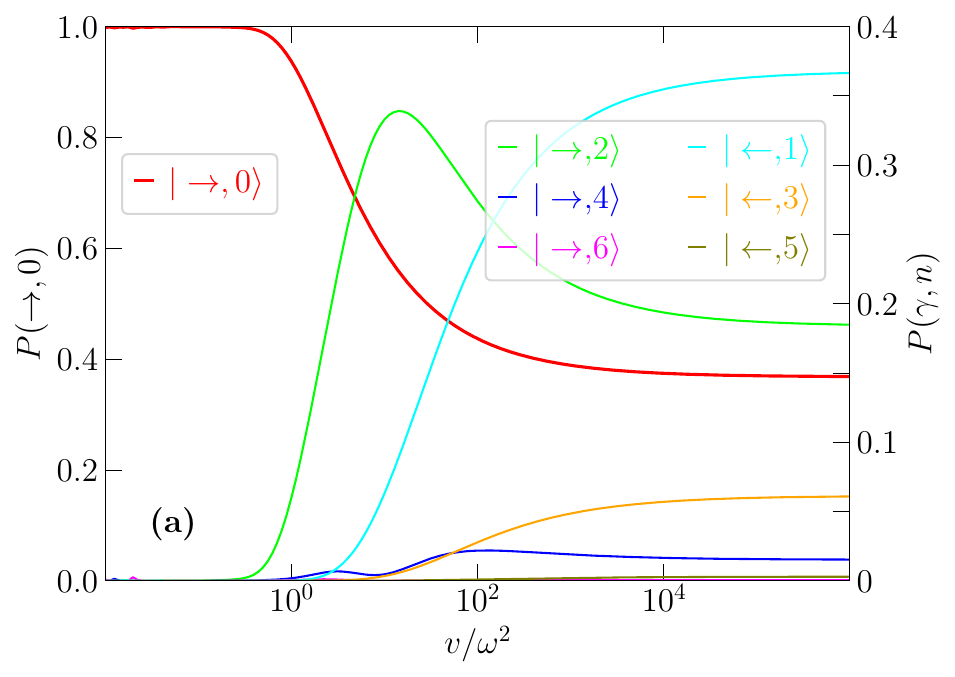}
\includegraphics[width=8.0cm]{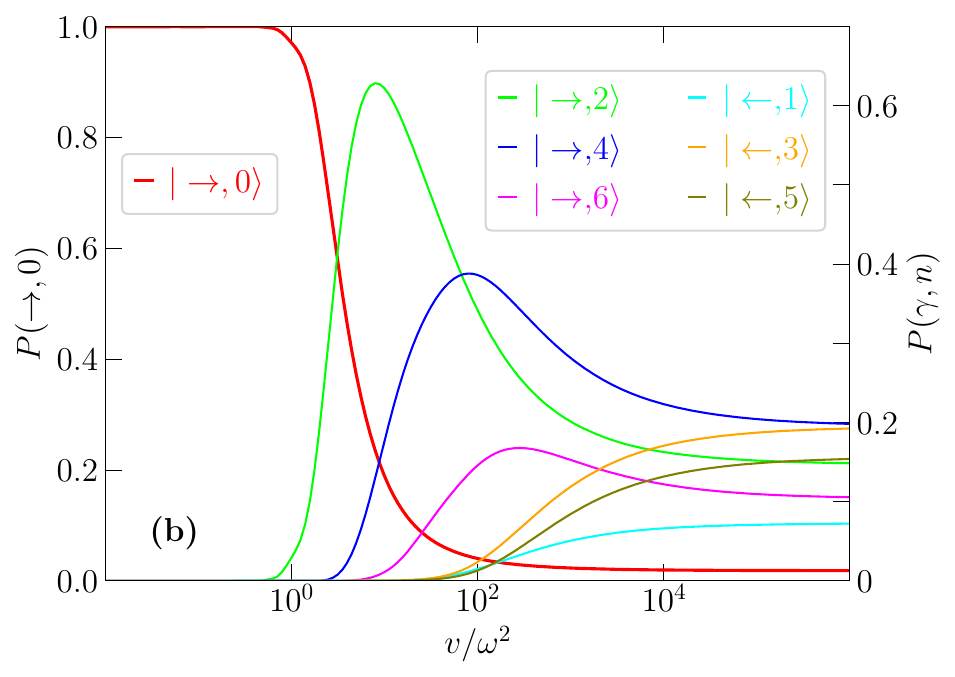}
\includegraphics[width=8.0cm]{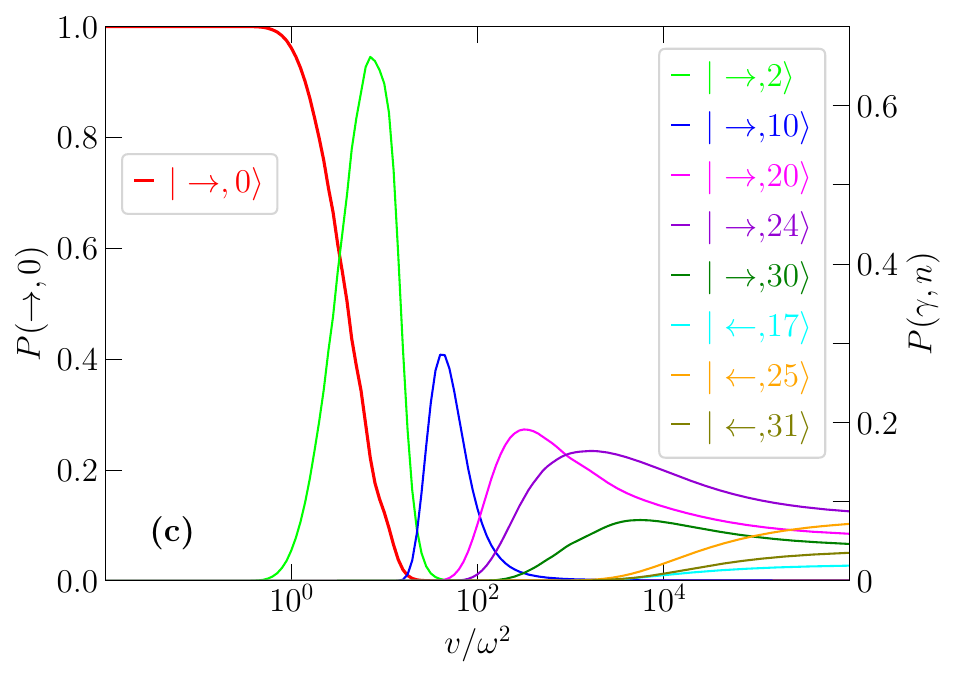}
\includegraphics[width=8.0cm]{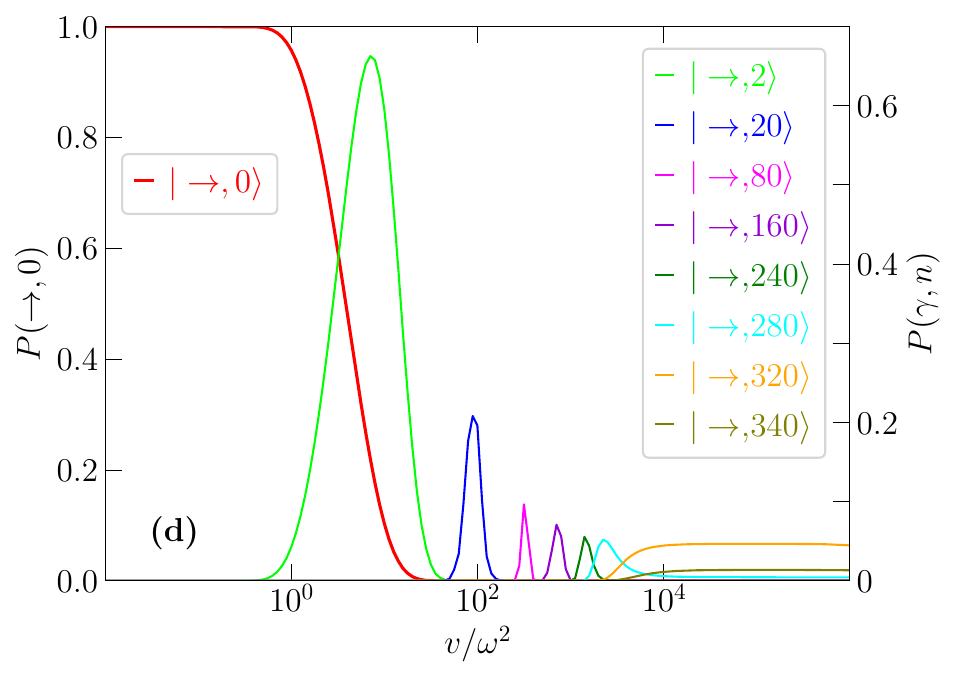}
\caption{Probabilities $P(\gamma,n)$ as functions of $v/\omega^2$ for the QRM following a sweep of $\Delta/\omega$ from zero to $4\times 10^3$, i.e.~from the strongly-correlated to the weakly-correlated regime. As in Fig.~\ref{Fig:RabiQuenchProbabilitiesNS}, $g/\omega=1$, 2, 5 and 20 in Panels (a), (b), (c) and (d), respectively. The y axis for $P(\rightarrow,0)$ is on the left-hand side of each frame, while the y axis for all other lines is on the right-hand side of the frame. The small fluctuations at small values of $v/\omega^2$ in Panel (a) are due to numerical errors that result from the finite size of the time steps in our simulations of the dynamics. These errors decrease with increasing $v/\omega^2$.}
\label{Fig:RabiQuenchProbabilitiesSN}
\end{figure}

\begin{figure}[h]
\includegraphics[width=8.0cm]{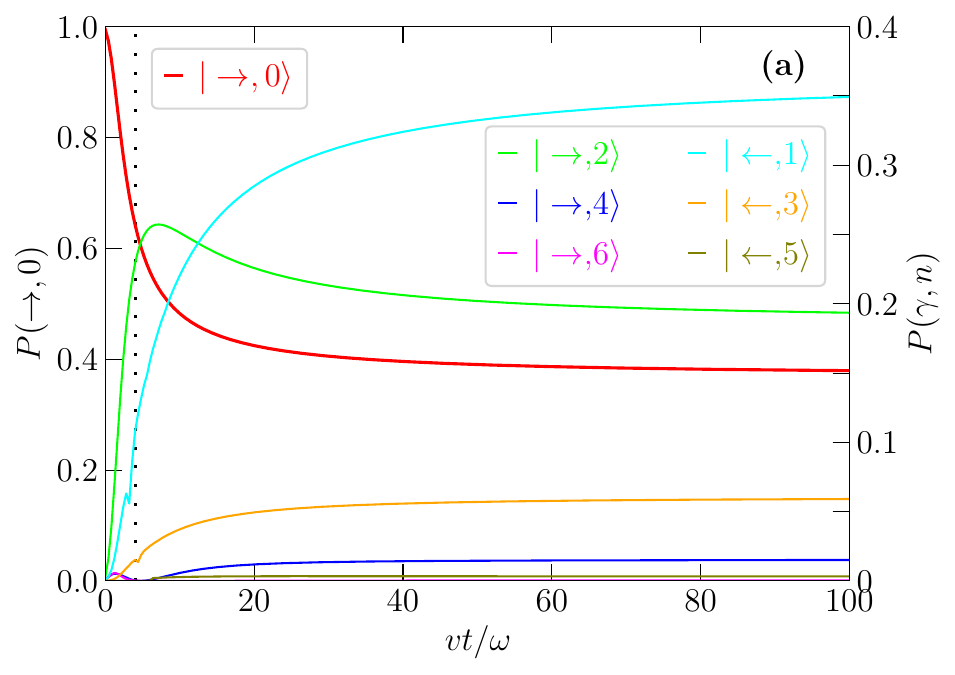}
\includegraphics[width=8.0cm]{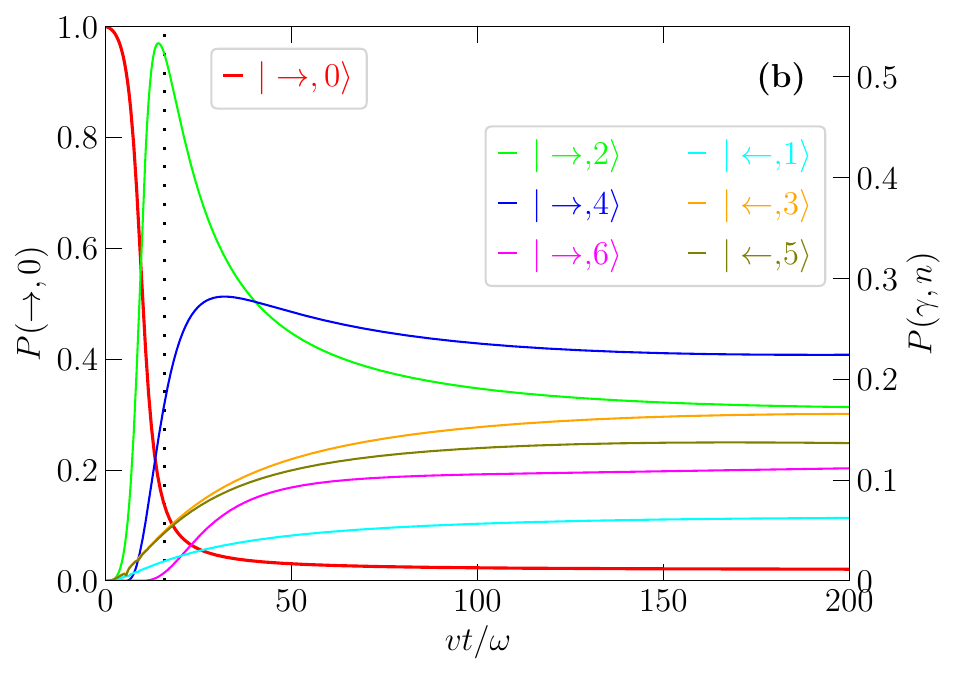}
\includegraphics[width=8.0cm]{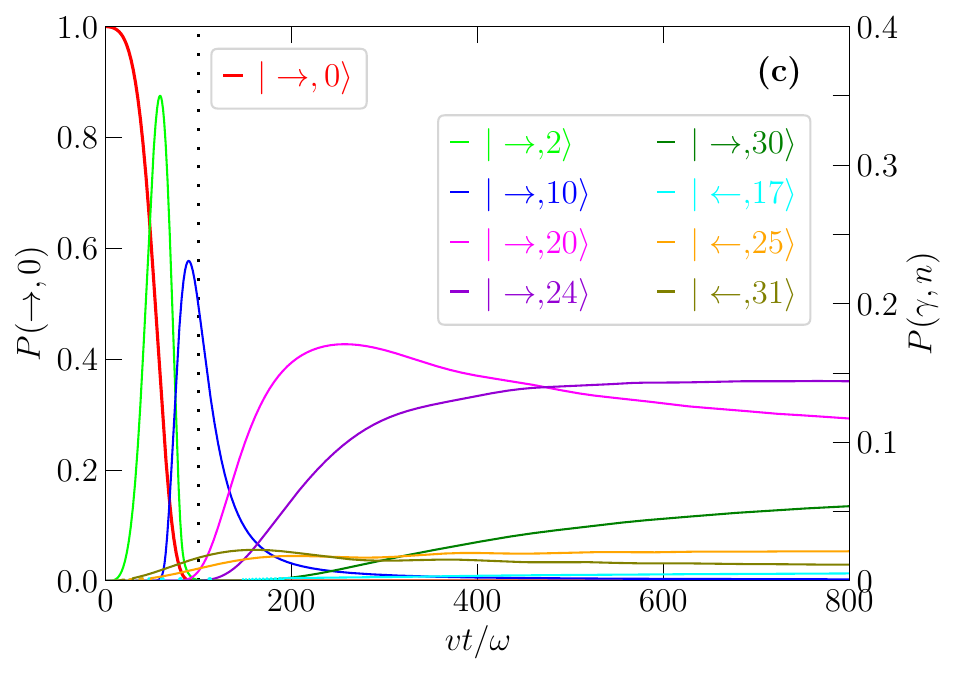}
\includegraphics[width=8.0cm]{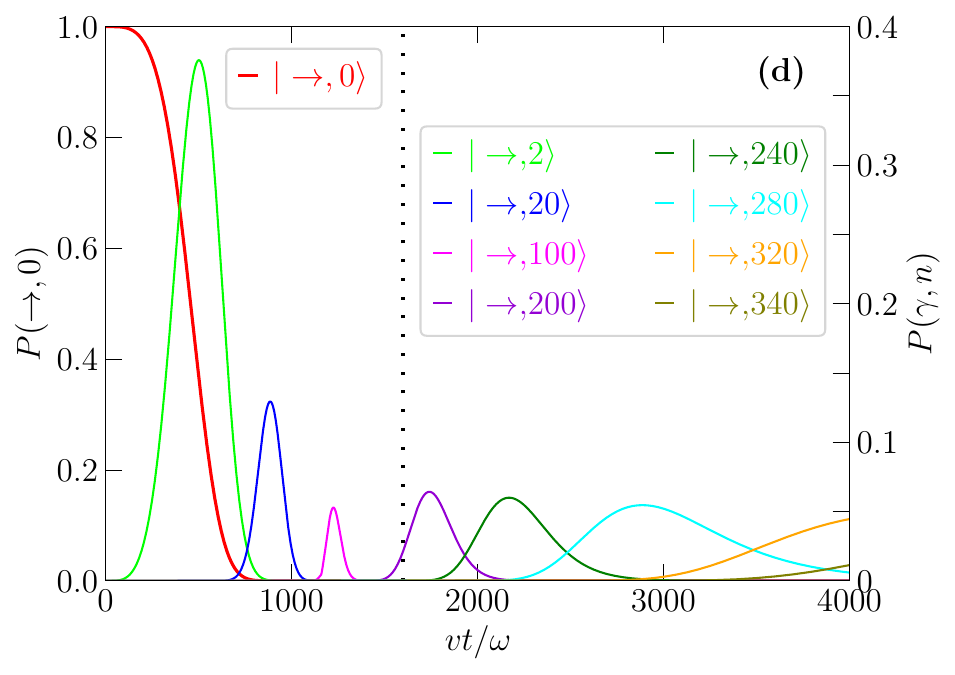}
\caption{Probabilities $P(\gamma,n)$ as functions of time $t$ for a strong-to-weak-correlation sweep. The four panels in this figure correspond to the four panels with the same labels in Fig.~\ref{Fig:RabiQuenchProbabilitiesSN}. In all cases, we set $v/\omega^2=10^4$. The vertical dotted lines mark the semiclassical phase transition crossing point. Although we label the different states by their asymptotic quantum numbers at $t\to\infty$, we calculate the probabilities using the instantaneous energy eigenstates at intermediate values of $t$. The small glitches most clearly seen in the cyan and orange lines around $vt/\omega=4$ in Panel (a) are due to avoided level crossings in the energy level structure, which complicate the definition of the different state probabilities.}
\label{Fig:RabiQuenchPopulationDynamicsSN}
\end{figure}

We now consider a sweep in the opposite direction, i.e.~with $\Delta_i/\omega=0$ and $\Delta_f/\omega=4\times 10^3$. We take the ground state in the parity-symmetric sector of the Hilbert space as the initial state. The results of the corresponding simulations are shown in Figs.~\ref{Fig:RabiQuenchProbabilitiesSN} and \ref{Fig:RabiQuenchPopulationDynamicsSN}. In these figures, the plotted quantities are the populations $P(\gamma,n)$ where $\gamma$ is $\rightarrow$ or $\leftarrow$. In the adiabatic limit, the system remains in its ground state, as expected. Away from the adiabatic limit, various excited states are populated with varying probabilities. In the fast-sweep limit $v/\omega^2\to\infty$, the probabilities are given by Eq.~(\ref{Eq:InitialFinalOverlap}), determined by the number of photons in the final state, independently of the qubit state. Since the qubit gap is the dominant energy scale at the final time, the system occupies highly excited states (corresponding to the qubit's excited state $\ket{\leftarrow}$) with about 50\% probability. The population of these highly excited states increases with increasing $g/\omega$. In the limit $g/\omega\gg 1$ and $v/\omega^2\to\infty$, the qubit ground-state and excited-state sectors are equally populated. Another feature that is different when we sweep from the strongly-correlated to the weakly-correlated regime is that the occupation probabilities of individual energy eigenstates reach higher values (which one can see by looking at the peaks of $P(\gamma,n)$ plotted as functions of $v/\omega^2$), especially in the case of large $g/\omega$ (Fig.~\ref{Fig:RabiQuenchProbabilitiesSN}c,d). From the point of view of state preparation, this situation allows a more deterministic preparation of specific energy eigenstates at the end of the sweep.

Examples of the dynamics of the probabilities $P(\gamma,n)$ as functions of time $t$ for the fast-sweep regime are shown in Fig.~\ref{Fig:RabiQuenchPopulationDynamicsSN}. For relatively small values of $g/\omega$ (Fig.~\ref{Fig:RabiQuenchPopulationDynamicsSN}a,b), the probability $P(\rightarrow,0)$ drops from its initial to its final value around the time $vt/\omega=(2g/\omega)^2$, similarly to what we saw in Fig.~\ref{Fig:RabiQuenchPopulationDynamicsNS}. For larger values of $g/\omega$ (Fig.~\ref{Fig:RabiQuenchPopulationDynamicsSN}c,d), $P(\rightarrow,0)$ drops to its final value (which is essentially zero) well before the semiclassical crossing point is reached. By the time that the semiclassical crossing point is reached, the system is already in a highly excited state. For all values of $g/\omega$, the different probabilities $P(\gamma,n)$ for low-lying final states exhibit transient peaks. However, unlike the behaviour shown in Fig.~\ref{Fig:RabiQuenchPopulationDynamicsNS}, the probabilities $P(\gamma,n)$ of higher energy levels rise monotonically and asymptotically approach their final values. As a result, the final state probabilities are relatively insensitive to the exact value of $\Delta_f$, in contrast to the dynamics shown in Fig.~\ref{Fig:RabiQuenchPopulationDynamicsNS}d. As in the case of $P(\gamma,n)$ as functions of $v/\omega^2$, the overall shapes of the curves is quite different when comparing the weak-to-strong and strong-to-weak-correlation sweeps.

One might wonder about the physical origin of the asymmetry between the two sweep directions. Since we take the initial state to be the ground state of the Hamiltonian at the initial time, in the weak-to-strong-correlation sweep, we effectively start with the ground state of a single well that evolves into a double well, while in the strong-to-weak-correlation sweep we start with the ground state of a double well that evolves into a single well. There is no reason to expect any symmetry between the two cases, except for the ground state probability. One crucial factor in the asymmetry between the two sweep directions pertains to the energy level order in the two limits $\Delta\to\infty$ and $\Delta\to 0$. In the limit $\Delta\to\infty$, the lowest energy levels correspond to the states $\ket{\rightarrow, 0}$, $\ket{\rightarrow, 2}$, $\ket{\rightarrow, 4}$, ..., with the states $\ket{\leftarrow, 1}$, $\ket{\leftarrow, 3}$, $\ket{\leftarrow, 5}$, ... having much higher energies. In contrast, when $\Delta\to 0$, the energy levels in increasing order correspond to the states $\ket{+,0}$, $\ket{-,1}$, $\ket{+,2}$, $\ket{-,3}$, ...

In relation to the normal-to-superradiant phase transition in the quantum Rabi model, it is worth mentioning the dynamical phase transition that occurs when the qubit or oscillator is driven on or near resonance \cite{Alsing}. At a certain driving amplitude the quasienergy spectrum collapses, and above this threshold amplitude no normalizable steady states exist. The phase transition in our case is different, because the spectrum collapses only at the transition point but simplifies away from the critical point on both sides of the transition. As such, we cannot use our results to infer the behaviour of a driven qubit-oscillator system as the one studied in Ref.~\cite{Alsing} with a linear sweep of parameters. It will be interesting to investigate a quench across the critical point in that situation in the future.

\section{Landau-Zener sweep of qubit coupled to oscillator}
\label{Sec:LZwO}

We now turn to the case of finite $\epsilon$ in Eq.~(\ref{Eq:QRM}), in particular sweeping $\epsilon$ from a large negative value to a large positive value:
\begin{equation}
\hat{H} = - \frac{\Delta}{2} \hat{\sigma}_x - \frac{vt}{2} \hat{\sigma}_z + \omega \hat{a}^{\dagger} \hat{a} + g \hat{\sigma}_z \left( \hat{a} + \hat{a}^{\dagger} \right).
\label{Eq:LZwOscillator}
\end{equation}
This situation is in fact close to models studied previously in the literature. The first two terms in Eq.~(\ref{Eq:LZwOscillator}) describe the LZ problem. The last two terms can be viewed from the perspective of the spin-boson model, in which the environment of a quantum two-level system is modeled as an infinite set of harmonic oscillators. As a result, combining the LZ problem with the spin-boson model to describe the LZ problem in a two-level system that is coupled to an environment gives a generalized version of Eq.~(\ref{Eq:LZwOscillator}) in which there are an infinite number of harmonic oscillators \cite{Ao,Kayanuma,Wubs,Nalbach}. The model described by Eq.~(\ref{Eq:LZwOscillator}), i.e.~with a single-mode environment, was also used to study the effect of a finite-temperature environment on the LZ problem \cite{Ashhab2014,Malla2018}. These past studies focused on the effect of the harmonic oscillators on the dynamics of the two-level system. Here we are equally interested in the state of the harmonic oscillator, which we view as part of the accessible and controllable quantum system, as opposed to being an unwanted environment that disrupts the dynamics of the controllable quantum system.

Instead of using the basis of bare states, i.e.~$\ket{\gamma, n}$ where the qubit state index $\gamma$ is $\uparrow$ or $\downarrow$ and the photon number operator $\hat{n}=\hat{a}^{\dagger} \hat{a}$, it is physically more meaningful to use a correlated basis in which the photon number $\hat{n}$ is defined as $\hat{n}_{\uparrow}=(\hat{a}^{\dagger}+g/\omega) (\hat{a}+g/\omega)$ for the qubit state $\uparrow$ and $\hat{n}_{\downarrow}=(\hat{a}^{\dagger}-g/\omega) (\hat{a}-g/\omega)$ for the qubit state $\downarrow$ \cite{Irish,Ashhab2010}. In other words, the states of the oscillator are displaced to account for the effective field induced by its interaction with the qubit. We use this modified basis below. It is worth noting that the displacement of the oscillator variables has the same physical origin as the displacement in the energy eigenstates in Eq.~(\ref{Eq:SuperradiantES}).

\begin{figure}[h]
\includegraphics[width=8.0cm]{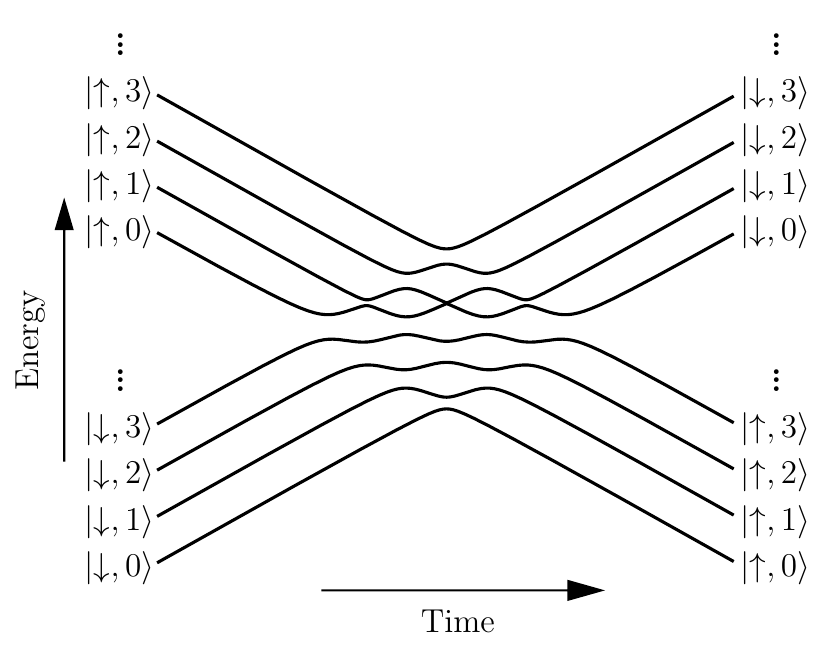}
\caption{Schematic diagram of energy levels in a problem of LZ sweep for a qubit coupled to a harmonic oscillator. Eight representative energy levels are shown. The energy level ladders extend to $\infty$, as indicated by the vertical dots.}
\label{Fig:LZwOEnergyLevels}
\end{figure}

To have an intuitive picture of the setup, we consider the case $\Delta=0$. This case leads to a simple energy level structure. As $t$ goes from $-\infty$ to $+\infty$, energy level crossings occur at the points $vt=m\omega$ with $m$ being any integer. At each value of $m$, the energy levels that correspond to the states $\ket{\downarrow, n}$ and $\ket{\uparrow, n+m}$ intersect, with $n=0, 1, 2, \cdots$. In other words, the energy level structure looks like a mesh containing an infinite number of energy level crossings. Taking a nonzero value of $\Delta$ turns all the energy level crossings into avoided crossings with gaps determined by the various system parameters. Figure \ref{Fig:LZwOEnergyLevels} shows a schematic diagram of the energy level structure, keeping only a few states that have the qubit in its state $\downarrow$ and a few states that have the qubit in its state $\uparrow$.

If the initial state at $t\to-\infty$ is the ground state, i.e.~$\ket{\downarrow, 0}$, the above intuitive picture gives a very good approximation for the final energy eigenstate probabilities, as long as $\omega$ is not much smaller than $\Delta$ \cite{Brundobler,Malla2021,Ashhab2023}. In other words, the probabilities $P(\gamma,n)$ are, to a very good approximation, given by the approximation that the different state populations are determined by a sequence of independent LZ transitions that occur as the system goes through the avoided crossings one by one:
\begin{eqnarray}
P(\uparrow, 0) & = & 1 - e^{-\pi\Delta_0^2/(2v)}
\nonumber \\
P(\uparrow, 1) & = & e^{-\pi\Delta_0^2/(2v)} \left( 1 - e^{-\pi\Delta_1^2/(2v)} \right)
\nonumber \\
P(\uparrow, 2) & = & e^{-\pi\Delta_0^2/(2v)} e^{-\pi\Delta_1^2/(2v)} \left( 1 - e^{-\pi\Delta_2^2/(2v)} \right)
\nonumber \\
& \vdots &
\nonumber \\
P(\downarrow, 0) & = & e^{-\pi \left( \sum_n \Delta_n^2 \right) / (2v)} = e^{-\pi\Delta^2/(2v)}
\nonumber \\
P(\downarrow, n) & = & 0 \hspace{1.0cm} {\rm for \ all} \ n\geq 1,
\label{Eq:LZwOProbabilities}
\end{eqnarray}
with
\begin{equation}
\Delta_n = \frac{1}{\sqrt{n!}} \left( \frac{2g}{\omega} \right)^n e^{-2(g/\omega)^2} \Delta.
\label{Eq:LZwOGaps}
\end{equation}
This expression for $\Delta_n$ resembles a Poisson distribution with $\left\langle n \right\rangle = (2g/\omega)^2$, because it is obtained by taking the overlap between the oscillator state $\ket{0}$ displaced by $g/\omega$ in one direction with the state $\ket{n}$ displaced by $g/\omega$ in the opposite direction \cite{Ashhab2010}. It is worth noting here that the expressions for $P(\downarrow, n)$ are exact \cite{Wubs}. Note also that the relation $\sum_n \Delta_n^2 = \Delta^2$ follows from the completeness of the photon-number basis states, and it ensures that the sum of all the probabilities in Eq.~(\ref{Eq:LZwOProbabilities}) is equal to one.

It can be intuitively expected that if the gaps at the different avoided crossings, i.e.~Eq.~(\ref{Eq:LZwOGaps}), are small relative to the energy scale separating the energy levels, i.e.~$\omega$, the picture of independent LZ transitions will be valid. This situation is realized when $\Delta/\omega\lesssim 1$ and/or $g/\omega\gg 1$. Outside these regimes, the energy level structure becomes more complex, and the LZ transitions cannot be considered as occurring independently. We will show the results of numerical simulations that test the validity conditions of Eq.~(\ref{Eq:LZwOProbabilities}). We note here that the extreme regime where $\Delta/\omega\gg 1$ and $g/\omega\gg 1$ is not of strong interest to us in this work, partly because both conditions  are difficult to realize experimentally, in addition to the fact that the corresponding simulations are computationally challenging. We therefore do not investigate this regime in detail in our numerical simulations.

\begin{figure}[h]
\includegraphics[width=8.0cm]{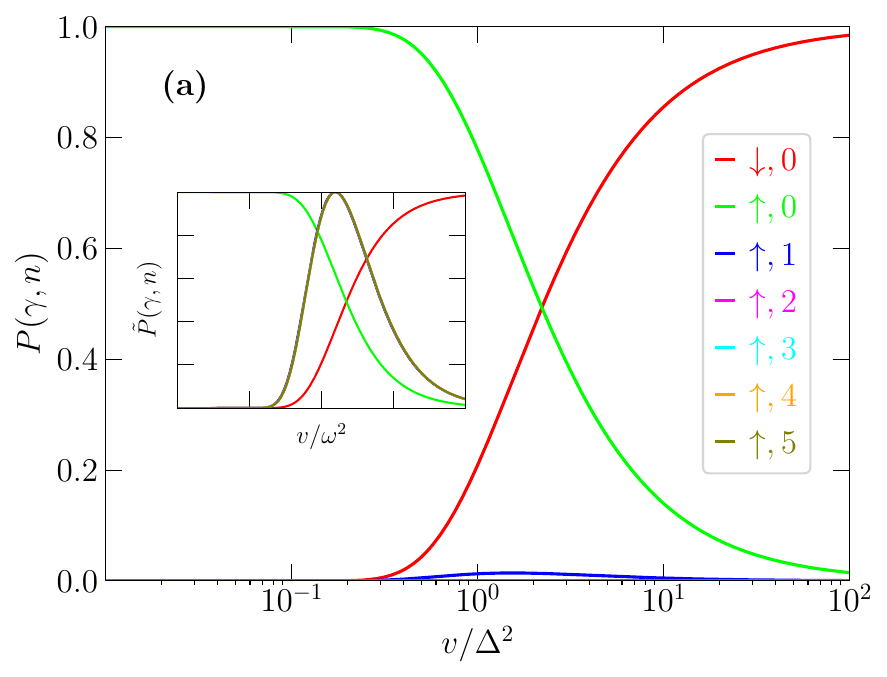}
\includegraphics[width=8.0cm]{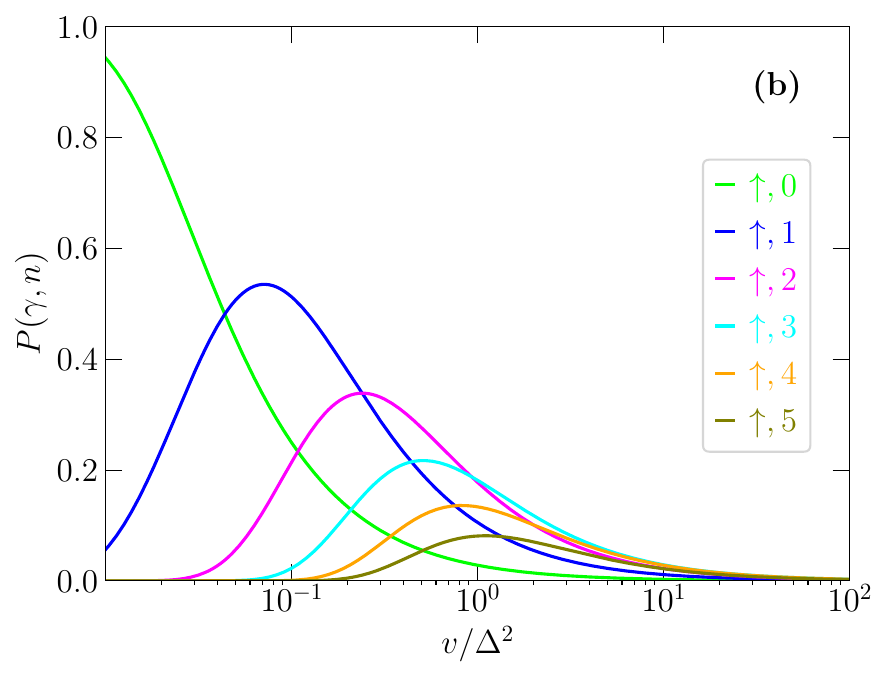}
\includegraphics[width=8.0cm]{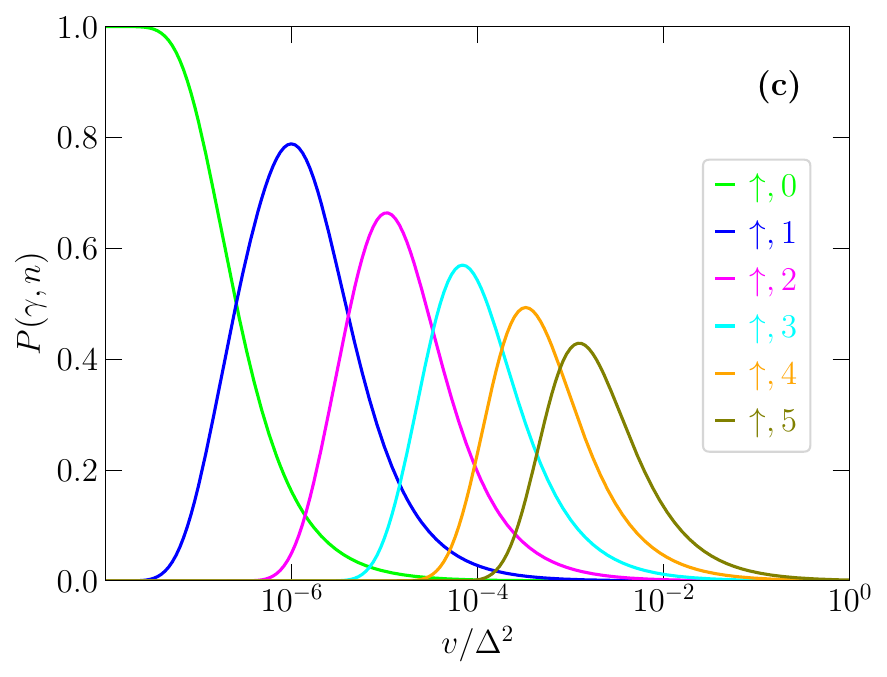}
\includegraphics[width=8.0cm]{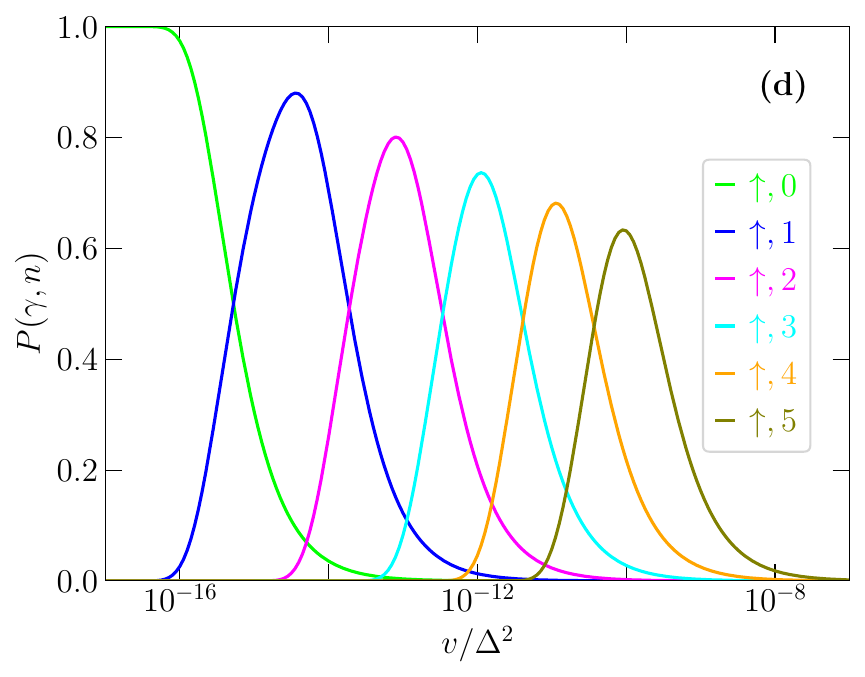}
\caption{Probabilities $P(\gamma, n)$ as functions of $v/\Delta^2$ for the LZ problem of a qubit coupled to a harmonic oscillator. The lines plotted in this figure are all based on Eq.~(\ref{Eq:LZwOProbabilities}), which is valid when $\Delta\lesssim\omega$. The red line corresponds to the state $\ket{\downarrow,0}$, whose final-time occupation probability $P(\downarrow,0)$ is independent of the coupling to the oscillator. The green, blue, magenta, cyan, orange and olive lines correspond to the states $\ket{\uparrow,n}$ with $n=0$, 1, 2, 3, 4 and 5, respectively. In Panels (a), (b), (c) and (d), $g/\omega=0.1$, 1, 2 and 3, respectively. The inset in Panel (a) shows the probabilities renormalized such that the maximum value of each line in the inset is one [$\tilde{P}(\gamma, n) = P(\gamma, n) / {\rm max}_{v} \{ P(\gamma, n) \}$], which reveals that the shapes of the curves $P(\uparrow, n)$ are essentially identical for $1 \leq n \leq 5$.}
\label{Fig:LZwO010}
\end{figure}

According to Eq.~(\ref{Eq:LZwOProbabilities}), the probabilities $P(\uparrow, n)$ depend on three independent parameters: $v/\Delta^2$, $g/\omega$ and $n$. In particular, the formulae suggest that the ratio $\Delta/\omega$ does not affect the structure of the probability peaks described by Eq.~(\ref{Eq:LZwOProbabilities}). The qubit gap $\Delta$ affects $P(\uparrow, n)$ only through the ratio $\Delta^2/v$. For example, an increase in $\Delta$ simply shifts all the $P(\gamma,n)$ curves to higher values of $v$. The formulae in Eq.~(\ref{Eq:LZwOProbabilities}) for a few of the representative states are plotted in Fig.~\ref{Fig:LZwO010}. As mentioned above, the probability for the qubit to remain in the initial state, i.e.~$\ket{\downarrow,0}$, is independent of the coupling to the oscillator. For the qubit state $\uparrow$, as one would intuitively expect, the occupation probability of the final-time ground state (i.e.~$\ket{\uparrow,0}$) approaches one in the adiabatic limit and approaches zero in the fast-sweep limit. Hence the plot of $P(\uparrow, 0)$ as a function of $v/\Delta^2$ looks generally like an inverted version of $P(\downarrow, 0)$. As $g/\omega$ increases, the curve of $P(\uparrow, 0)$ shifts to the left, i.e.~the drop of $P(\uparrow, 0)$ from one to zero occurs at smaller values of $v/\Delta^2$.

All $P(\uparrow, n)$ plots with $n\geq 1$ are simple peaks at intermediate values of $v/\Delta^2$, with asymptotic values of zero in both the adiabatic and fast-sweep limits. The locations and heights of the peaks depend on the qubit-oscillator coupling in rather nontrivial ways. In the limit of small $g/\omega$ (Fig.~\ref{Fig:LZwO010}a), the shapes of the $P(\uparrow, n)$ peaks are extremely weakly dependent on $n$, except for an overall reduction in the scale with increasing $n$; specifically
\begin{equation}
\frac{P(\uparrow, n)}{P(\uparrow, n-1)} \approx \frac{(2g/\omega)^2}{n}.
\end{equation}
The reason behind this simple relation is that all the factors in the formula for $P(\uparrow, n)$ in Eq.~(\ref{Eq:LZwOProbabilities}) except for the first and last factors, i.e.~the product $e^{-\pi\Delta_0^2/(2v)} \left( 1 - e^{-\pi\Delta_n^2/(2v)} \right)$, are approximately equal to one at the location of the peak, and the last factor can be approximated as $\left( 1 - e^{-\pi\Delta_n^2/(2v)} \right) \approx \pi\Delta_n^2/(2v) \propto (2g/\omega)^{2n}/n!$ at the location of the peak. This result means, among other things, that one does not have much control over the number of created photons by controlling the sweep rate $v$. In fact, the probability of creating any photons in the oscillator remains small for all values of $v/\Delta^2$.

In the case where $g/\omega$ is not much smaller than one (Fig.~\ref{Fig:LZwO010}b-d), the probabilities $P(\uparrow, n)$ have more structure. Comparing the plots of $P(\uparrow, n)$ as functions of $v/\Delta^2$ for different values of $n$, the peak locations shift to higher values of $v/\Delta^2$ and decrease in height as $n$ increases. As $g/\omega$ increases, we obtain more separation between the different peaks, in addition to having higher peaks. As a result, for large $g/\omega$ (Fig.~\ref{Fig:LZwO010}d), each $P(\uparrow, n)$ peak reaches a height of almost one. This result means that by choosing the appropriate value of $v/\Delta^2$, one can deterministically prepare a Fock state with a specific value of $n$. This result can be understood intuitively as follows: the gaps in the different avoided crossings (Eq.~\ref{Eq:LZwOGaps}) scale as $(2g/\omega)^n/\sqrt{n!}$. For $g/\omega\gg \sqrt{n}$, the gaps obey the relation $\Delta_n\gg\Delta_{n-1}$. If we choose $v$ such that $\Delta_n^2 \gg v \gg \Delta_{n-1}^2$, the first $n$ avoided crossings (from $\ket{\uparrow, 0}$ to $\ket{\uparrow, n-1}$) will be traversed fast, keeping the system in the state $\ket{\downarrow, 0}$, while the avoided crossing with the state $\ket{\uparrow, n}$ is traversed adiabatically, transferring the population to the state $\ket{\uparrow, n}$. It should be noted that to achieve the full separation between the probability peaks up to photon number $n$, it is necessary to have $(g/\omega)^2/n\gtrsim 10$, which is based on the fact that the LZ probability goes from almost zero to almost one or vice versa in a $v/\Delta_n^2$ range of about one order of magnitude (see e.g.~Fig.~\ref{Fig:LZwO010}a). This requirement can make the gap $\Delta_n$ extremely small because of the factor $e^{-2(g/\omega)^2}$ and in turn require a proportionately slow sweep. As a result, it is extremely difficult to implement this scenario in a realistic experimental setup, as we will discuss in Sec.~\ref{Sec:Experiment}.

\begin{figure}[h]
\includegraphics[width=16.0cm]{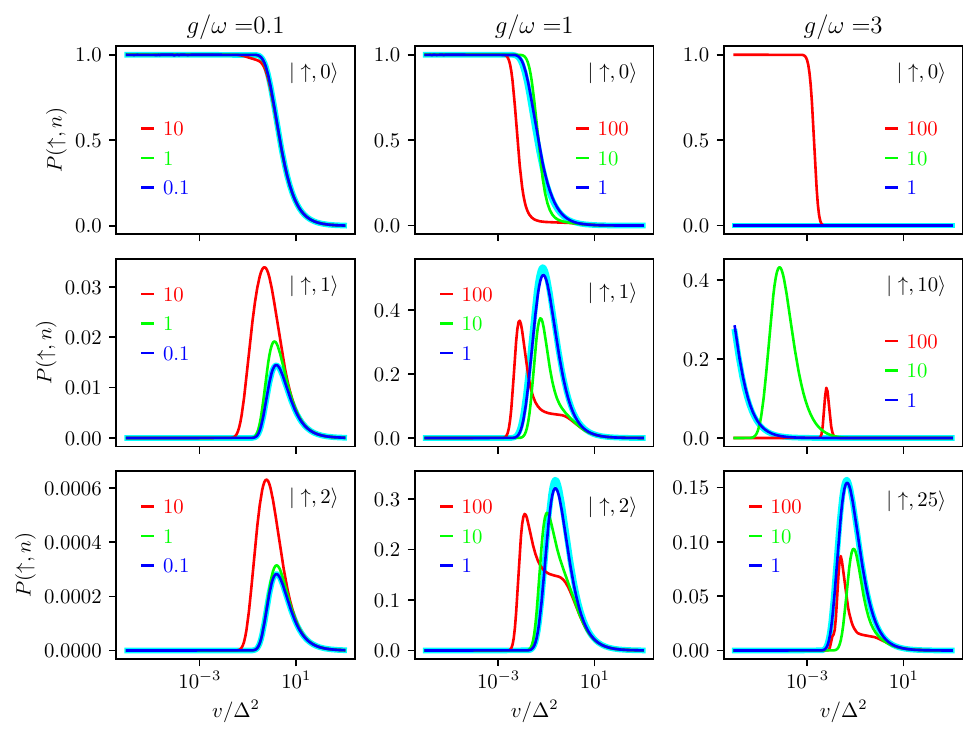}
\caption{Probabilities $P(\uparrow, n)$ as functions of $v/\Delta^2$ for the LZ problem of a qubit coupled to a harmonic oscillator. The thick cyan lines in all panels are obtained using Eq.~(\ref{Eq:LZwOProbabilities}), while all other lines show the results of numerical simulations of the dynamics. In the left column, we set $g/\omega=0.1$. In the middle column, we set $g/\omega=1$. In the right column, we set $g/\omega=3$.  Within each column, the different rows correspond to different photon numbers in the final state, as indicated at the top-right corner of each panel. Different lines in each panel correspond to different values of $\Delta/\omega$; the values of $\Delta/\omega$ are indicated in the legends. The blue line coincides almost perfectly with the cyan line in all the panels, which is what we expect for small values of $\Delta/\omega$. A variety of deviations appear for larger values of $\Delta/\omega$.}
\label{Fig:LZwOTable}
\end{figure}

We now examine the validity conditions for Eq.~(\ref{Eq:LZwOProbabilities}). We performed numerical simulations of the dynamics with a variety of parameters. We varied simulation parameters such as the size of the Hilbert space and the initial and final times to ensure that we obtain well-converged results, i.e.~results that are not significantly distorted by finite-size effects in the simulations. In particular, since we do not make approximations such as the weak-coupling and/or rotating-wave approximations, the results of the numerical simulations should be valid for small or large qubit-oscillator detuning and for weak or strong coupling. The results are plotted in Fig.~\ref{Fig:LZwOTable}. As expected, when $\Delta/\omega\lesssim 1$, the numerical simulation results agree with the predictions based on Eq.~(\ref{Eq:LZwOProbabilities}), because all the avoided crossings are well separated from each other. When $\Delta/\omega\gg 1$, Eq.~(\ref{Eq:LZwOProbabilities}) is no longer valid. In this case, the gaps given by Eq.~(\ref{Eq:LZwOGaps}) are in general larger than the distances between the avoided crossings, such that we can no longer think of the dynamics as a sequence of independent probability redistribution processes. We therefore obtain large deviations between the simulation results and Eq.~(\ref{Eq:LZwOProbabilities}). A few differences are worth highlighting here. In the case of large $g/\omega$, Eq.~(\ref{Eq:LZwOProbabilities}) predicts that an extremely slow sweep is required to remain in the ground state. The case with $g/\omega=3$ and $\Delta/\omega=100$ shows that this prediction fails badly when the energy level structure becomes more complex than the simple mesh assumed in deriving Eq.~(\ref{Eq:LZwOProbabilities}). Another difference is that some of the probabilities have peaks with side shoulders, rather than the single peaks described by Eq.~(\ref{Eq:LZwOProbabilities}). We suspect that this dependence arises when more than two energy levels are significantly populated during the system's evolution, which can lead to quantum interference effects and nontrivial peak shapes. We finally point out that the progression of the curves as we increase $\Delta/\omega$ is not unidirectional, as can be seen in the bottom-right panel, where the center of the peak moves to the right and then to the left as we go from $\Delta/\omega=1$ to $\Delta/\omega=10$ to $\Delta/\omega=100$.

So far, we have considered the case of a qubit coupled to a single oscillator. The problem can be generalized to the case of a qubit coupled to multiple oscillators, or alternatively a multi-mode resonator \cite{Stehli}. The Hamiltonian is then given by
\begin{equation}
\hat{H} = - \frac{\Delta}{2} \hat{\sigma}_x - \frac{vt}{2} \hat{\sigma}_z + \sum_j \left[ \omega_j \hat{a}_j^{\dagger} \hat{a}_j + g_j \hat{\sigma}_z \left( \hat{a}_j + \hat{a}_j^{\dagger} \right) \right].
\label{Eq:LZwMultipleOscillators}
\end{equation}
As with the case of a single oscillator, the energy level associated with the state $\ket{\downarrow,0,0,0,\cdots}$ encounters an infinite sequence of avoided crossings. The order at which these avoided crossings are encountered depends on the energies of the oscillators. Specifically, the crossing with the state $\ket{\uparrow,n_1,n_2,n_3,\cdots}$ occurs at the point $vt = \sum_j n_j \omega_j$. Upon making the polaron transformation for the harmonic oscillators, we obtain the effective gaps of the avoided crossings:
\begin{equation}
\Delta_{n_1,n_2,n_3,\cdots} =  \Delta \times \prod_j \frac{1}{\sqrt{n_j!}} \left( \frac{2g_j}{\omega_j} \right)^{n_j} e^{-2(g_j/\omega_j)^2}.
\label{Eq:LZwMOsGaps}
\end{equation}
If the relevant avoided crossings are well separated from each other, the occupation probabilities can be approximately evaluated by calculating how much of the probability is transferred to the states $\ket{\uparrow,n_1,n_2,n_3,\cdots}$ as the avoided crossings are traversed one by one. In particular, in the case when the largest avoided-crossing gaps correspond to excited states in multiple oscillators that are deep-strongly coupled to the qubit, with $g_j/\omega_j\gg 1$, at least some of the states $\ket{\uparrow,n_1,n_2,n_3,\cdots}$ can be prepared almost deterministically by choosing a sweep rate that is adiabatic for the relevant avoided crossing but fast for all the avoided crossings that are encountered before it. This scenario obviously requires that all the gaps that are encountered earlier in the sweep must be much smaller than the one associated with the target state.

Another interesting question in this context is related to the case where one harmonic oscillator is coupled strongly to the qubit while all other oscillators are coupled weakly. In this case, we effectively obtain a quantum system that comprises the qubit and the strongly coupled oscillator, while all the weakly coupled oscillators serve as an environment for the quantum dynamics. The physics of LZ transitions in an open multilevel system is not very well understood \cite{Ashhab2016}. However, we can make some general statements about this case. In particular, since the environment couples to the operator $\hat{\sigma}_z$, one effect of the environment is to suppress quantum interference between multiple LZ transitions. As such, the environment pushes the final state probabilities to the expressions given in Eq.~(\ref{Eq:LZwOProbabilities}).

\section{Experimental considerations for superconducting circuits}
\label{Sec:Experiment}

In the previous sections we performed systematic simulations of various scenarios and looked for theoretically interesting features in the simulation results. We now make a link between our results and possible implementations using superconducting circuits. Superconducting resonators typically have frequencies $\omega$ ranging from hundreds of MHz to a few GHz. Qubit gaps $\Delta$ are typically in the range 1-10 GHz. In some qubit designs, the gap is tunable and can be tuned down to zero in principle. However, practical considerations such as decoherence make any experimental measurement practically impossible below a few hundred MHz. The bias $\epsilon$ can be tuned between negative and positive values of tens of GHz. Qubit-resonator coupling strengths are typically tens or hundreds of MHz. However, GHz-scale coupling strengths have also been achieved.

The record value of $g/\omega$ is 1.34, reported in Ref.~\cite{YoshiharaNP}. For both the observation of Kibble-Zurek-like excitation generation in a quench between the normal and superradiant phases investigated in Sec.~\ref{Sec:Quench} and the deterministic generation of multi-photon Fock states investigated in Sec.~\ref{Sec:LZwO}, larger values of $g/\omega$ are needed. The extreme values of $g/\omega$ needed for these purposes could be engineered dynamically by modulating the system parameters to obtain radically different effective system parameters \cite{Ballester,Langford,Qin,Leroux}. In particular, by applying a parametric drive to the resonator frequency, the ratio $g/\omega$ can be enhanced to an effective value that is a few orders of magnitude larger than that of the undriven system with relatively moderate drive parameters \cite{Leroux}. The $g/\omega$ values used in all the panels in Figs.~\ref{Fig:RabiQuenchProbabilitiesNS}-\ref{Fig:LZwOTable} can be achieved in a suitably rotating frame. The sweep rates required for observing excitation generation in a normal-superradiant or superradiant-normal quench are rather moderate and should be achievable in such a dynamical realization of the effective Hamiltonian. The deterministic generation of multi-photon Fock states (Fig.~\ref{Fig:LZwO010}d) would require $v$ values well below 1 MHz per second, even using the lab frame value of $\Delta\sim 10$ GHz. As such, realizing this phenomenon in superconducting circuits using a simple sweep of $\epsilon$ or resonator frequency modulation as described in Ref.~\cite{Leroux} seems unrealistic. It is possible that this phenomenon might be observable using alternative techniques or different physical systems. It is also worth noting that if we have a highly controllable and stable quantum system, it could be possible to use a nonlinear sweep in which the sweep rate is tuned in real time to be adiabatic or fast for different avoided crossings based on the intended purpose of the experiment.

\section{Conclusion}
\label{Sec:Conclusion}

We have investigated the dynamics and in particular the final states of qubit-oscillator systems following a finite-duration quench in which one of the system parameters is swept between two values that correspond to qualitatively different states of the system. We obtained a number of results, some of which add theoretical insight to our understanding of quenched quantum system dynamics, and some of which are relevant to possible experimental realizations for practical applications. In our analysis, we paid special attention to the possibility of using parameter sweeps to generate states of interest for quantum information processing tasks.

In one case we investigated a sweep from the normal to the superradiant phase or vice versa. As expected, we found that excitations are created as a result of the quench between the two different phases. While the presence or absence of a sharp phase transition did not manifest itself in the final state probabilities, it led to qualitative differences in the dynamics of the quenched system. These results improve our understanding of quantum quenches through a phase transition point, especially as they relate to a rather unusual phase transition.

In another case we investigated an LZ sweep of a qubit coupled to an oscillator. In this case we found a variety of behaviours depending on the relation between the system parameters, including a case of extreme sensitivity in the final state to the sweep rate. From the point of view of quantum state preparation, a particularly interesting result that we found is the ability to almost deterministically prepare states with a specific number of excitation quanta in the oscillator. Although the peaks are narrow, requiring a fine tuning of the sweep rate, this protocol has the advantage of being robust against static fluctuations that shift the initial and final values of the qubit bias.

Our analysis also sheds light on the topic of excitation creation in a harmonic oscillator via an LZ sweep in a coupled qubit, which has not been investigated in the literature. We analyzed the predictions of a theoretical formula based on the assumption of independent LZ transitions and established the regime of validity of the theoretical formula.

In this work we focused on a single, linear sweep of parameters. Linear sweeps can also be repeated to create a situation of periodic driving \cite{Oliver,SaitoShiro,Sillanpaa,Zhou,Shevchenko}, which is one of the standard approaches to generating specific target states in quantum systems. Our results can form the basis for new protocols for state preparation in cavity-QED systems for quantum computing, communication and sensing applications.

\section*{Acknowledgment}

We would like to thank Neill Lambert, Keiji Saito and Sergey Shevchenko for useful discussions. This work was supported by Japan's MEXT Quantum Leap Flagship Program Grant Number JPMXS0120319794 and by Japan Science and Technology Agency Core Research for Evolutionary Science and Technology Grant Number JPMJCR1775.

\end{document}